\documentclass[conference]{IEEEtran}

\usepackage{url}
\usepackage{graphicx}

\usepackage{amsmath}

\newcommand{\TTL}{{\tt TTL} }
\newcommand{\IPid}{{\tt IPid} }

\begin{document}

\makeatletter \renewcommand\@biblabel[1]{$^{#1}$} \makeatother
\thispagestyle{empty}
\title{Forecasting Full-Path Network Congestion Using One Bit
Signalling}
\author{\IEEEauthorblockN{Mussie Woldeselassie}
\IEEEauthorblockA{Department of E\&EE\\University College London\\
Torrington Place\\
London WC1E 7JE, UK\\
Email: mwoldese@ee.ucl.ac.uk}
\and
\IEEEauthorblockN{Richard G. Clegg}
\IEEEauthorblockA{Department of E\&EE\\University College London\\
Torrington Place\\
London WC1E 7JE, UK\\
Email: richard@richardclegg.org}
\and
\IEEEauthorblockN{Miguel Rio}
\IEEEauthorblockA{Department of E\&EE\\University College London\\
Torrington Place\\
London WC1E 7JE, UK\\
Email: m.rio@ee.ucl.ac.uk}
}
\maketitle

\thispagestyle{headings}
\begin{abstract}
In this paper, we propose a mechanism for packet marking called
Probabilistic Congestion Notification (PCN). This scheme makes use of
the 1-bit
Explicit Congestion Notification
(ECN) field in the Internet Protocol (IP) header.  It allows the
source to estimate the exact level of congestion at each intermediate
queue. By knowing this, the source could take avoiding action either
by adapting its sending rate or by using alternate routes. The
estimation mechanism makes use of time series analysis both to improve
the quality of the congestion estimation and to predict, ahead of
time, the congestion level which subsequent packets will encounter.

The proposed protocol is tested in ns-2 simulator using a background
of real Internet traffic traces. Results show that the methods can
successfully calculate the congestion at any queue along the path with
low error levels.

\end{abstract}



\section{Introduction}
\label{sec:intro}

Today's packet based Internet Protocol (IP)
relies on the congestion control in
Transmission Control Protocol
(TCP) for stability \cite{j88con}. TCP congestion mechanisms react to
congestion by adjusting a ``congestion window" according to whether
packets are received or lost. Many variants of TCP have been proposed
which alter either the means by which congestion is detected or the
response to that congestion. The key contribution of this paper is
twofold.  Firstly, a new protocol Probabilistic Congestion
Notification (PCN) is proposed for probabilistic packet marking which
uses only a single bit,
the Explicit Congestion Notification
(ECN) bit in the IP header  \cite{ECN}. In PCN, routers do not maintain
any per-flow state of the flows. The scheme allows an end host to
estimate not only the level of congestion that packets encountered but
also to track the levels of congestion separately for each of the
intervening routers.
Congestion information from all routers will be useful in several
scenarios like multi-path TCP, overlay peer selection or future user
selectable routing schemes, for
further motivation see \cite{XCP}.  At its simplest, taking
the largest such estimate along the path
can be used to estimate the bottleneck
link.

Secondly, time series analysis methods (see Box-Jenkins \cite{GBox}
further details) are used to both improve the accuracy of this
estimate and to allow the level of congestion to be predicted. This
allows end hosts to decide on their actions based upon the level of
congestion a packet {\em will encounter\/} if it is sent now rather
than the level it {\em  would have encountered\/} had it been sent at
some previous time. 

In order to test the protocol and prediction real traffic traces have
been used to provide realistic background traffic.  The results show
that PCN
can produce a reliable estimate for the true congestion level on all
routers with low root mean square error and bias.  Without the time
series analysis correction, PCN has much higher errors unless long
sampling times are used.  The scheme described predicts the level of
congestion at all
routers between two end hosts.

\subsection{Background}
\label{sec:background}


Router based packet marking schemes are based upon one common idea,
that routers notify end hosts of congestion by modifying one (or more)
bits in the packet header \cite{RFC3168}. Examples include the Random
Exponential  Marking (REM) \cite{REM}, Random Additive  Marking (RAM)
\cite{RAM}, Deterministic Quantisation Marking (DQM) \cite{DQM},
Variable-structure Congestion-control Protocol (VCP) \cite{VCP} and
eXplicit Congestion Protocol (XCP) \cite{XCP}. The schemes can be
differentiated from each other by whether the marking is deterministic
or probabilistic, by how many bits in the header they use and by
whether they attempt to calculate congestion on the whole path or on
individual
routers along a path
(see section \ref{sec:comparison} for further detail).

Different schemes vary how the level of congestion is estimated. In
one common scheme (used by VCP amongst others) the router estimates
its level of congestion using the {\em load factor\/}. The load factor
is described in \cite{ATMRJSK}, it is an estimate of current local
congestion which is tracked by every router that is capable of the
packet marking scheme. The load factor is estimated for each outgoing
link for intervals of size $t_{\rho}$.  The value $t_{\rho}$ should be
larger than the round-trip time (RTT) of most flows, but small enough
to capture the dynamic changes in traffic level \cite{VCP}.   In this
paper $t_{\rho}$ of 200ms is used
since this is the value suggested in \cite{VCP}.
The load factor within period $l$ is given by
\begin{equation}
\label{eqn:pcnlf}
\rho_{l} =
\frac{\lambda_{l}+\kappa_{q}{\hat{q_{l}}}}{\gamma_l{C_l}{t_{\rho}}},
\end{equation}
where, $\lambda_{l}$ is the amount of input traffic (number of
packets) during period $l$, $\hat{q_l}$ is the persistent queue length
during period $l$ (measured with a low pass filter),  $\kappa_q$
controls how fast the persistent queue drains, $\gamma_l$ is the
target utilisation (set to a value close to 1) and $C_l$ is outgoing
link capacity.  For further
details of the precise formulation of these quantities see
\cite{ATMRJSK}.

\section{The proposed packet marking scheme}
\label{sec:marking}
This section describes the new packet marking scheme, probabilistic
congestion notification (PCN).  This scheme
uses only a single bit, does not require per-flow state at the router
and produces an estimate
for the congestion level at each queue the traffic encounters on its
outward
journey.

\subsection{Protocol at router and end hosts}
\label{sec:protocol}

The PCN scheme marks a single bit in the packet header, this is used
with statistical methods to estimate the congestion of any of several
possible intermediate routers.  PCN relies on two fields in
the IP header: the IP identifier (IPid) and the time to live (TTL)
fields.
These are initialised by the original source of the packet
and the TTL field is reduced by one at every hop on the path.

The marking scheme at the router is both stateless and extremely
simple.
Assume each intermediate router has a current load factor
(\ref{eqn:pcnlf}) which the source wishes to estimate. The load factor
($\rho_l$) is in the range  [0,100]
(values above 100 are rounded down).  This range is not a necessary
assumption of
the algorithm as discussed later.

Assume that there are at most $M$ intervening routers 
-- initially this is fixed at 32 for all connections.
It is not a problem for the algorithm if there are fewer intervening
routers (but it does cause loss of efficiency).  If there are more
then some information will be split between routers
(see later
discussion).
The strategy
is to allow each outgoing packet to have its ECN bit set by at most
one of
the intervening routers.  This ECN bit is set in a probabilistic
manner
governed by the load factor in such a way that the ratio of marked
packets
to total packets is equal to the load factor (divided by 100).
Whether the ECN bit is marked or not marked
is communicated back to the source on the acknowledgement (ACK) for
the outgoing packet
(so the receiver simply has to copy the state
of the ECN bit onto the ACK).  The IPid and TTL are used to determine
which router can mark the packet.

Consider the condition
\begin{equation}
\TTL \mod M = \IPid \mod M.
\label{eqn:ttlcond}
\end{equation}
By
definition the IPid remains constant and
TTL decreases by one at each hop.  Therefore,
if there are at most $M$ intervening routers then for only one such
router will this condition be true.  Only the router for which
(\ref{eqn:ttlcond}) is true may set the ECN bit on a packet.
Define a packet as {\em markable\/} by router $i$ if
(\ref{eqn:ttlcond}) is true for that packet for router $i$.
The source knows the IPid and TTL for every packet and
can calculate for which router a given packet was markable.

The router for which condition (\ref{eqn:ttlcond}) is met marks
the packet with a probability equal to the load factor.
The expectation value of the proportion of markable packets with
the ECN bit set by router $i$ is equal to its load factor (while
the load factor remains constant -- see section \ref{sec:prediction}
for more discussion of this issue).
Therefore, the proportion of ACKs from packets
markable by router $i$ which have their ECN
bit set is an unbiased estimate of the load factor.

The full protocol for PCN can be simply given as follows.
\begin{itemize}
\item The source sets the ECN bit to zero, TTL sets to $M$, and
increments the IPid by one
for each packet.
\item Intermediate routers, if condition (\ref{eqn:ttlcond}) is met,
set
the ECN bit with a probability equal to their load.
\item The receiver copies the ECN bit from a packet onto that packet's
ACK (there must be one ACK for every packet).
\item The source tracks the ECN bits on ACKs to estimate congestion
on intermediate routers.
\end{itemize}

This algorithm replaced ECN at routers and cannot coexist with it
(routers performing standard ECN marking will confuse PCN estimates).
It is this final part which enables the source to estimate congestion
and which will be a main focus of the results in this paper.  Section
\ref{sec:prediction} shows how time series modelling techniques can
be used to improve the estimate of congestion level and section
\ref{sec:results} shows ns-2 modelling results which prove the scheme
practical for realistic estimation scenarios.

The value $M = 32$ was chosen since \cite{hopCount}
shows that a hop count of more than 30 is extremely rare in the
real Internet.  However, if there are more than $M$ routers the
protocol's failure mode is not a major issue although some packets may
be marked by more than one router.

If there are less than $M$ intervening routers some packets
are not markable by any and an
opportunity to get data is lost.
A possible improvement is to pre-signal the actual
number of intervening routers (by communicating the TTL of the SYN
packet).  The PCN source can ensure that only packets
which have condition (\ref{eqn:ttlcond}) met for some intervening
router are sent.  This will increase the number of samples at each
router.
This improvement is also tested in section \ref{sec:results}.

Estimates of congestion other than that given by (\ref{eqn:pcnlf}) can
easily by used by PCN.  If the new load equation is
not in the range $[0,100]$ then a simple linear transform, $al+b$,
will map
it into this range.  If some regions of the range are more of interest
than others then nonlinear transforms could be used.

\subsection{Comparison with other packet marking schemes}
\label{sec:comparison}

It is important to recognise how PCN differs from other packet marking
schemes recently suggested in the literature.
Like XCP, PCN estimates the congestion level at each intervening
queue.  However,
XCP requires a new 128 bit header to record the information whereas
PCN requires
only a single ECN bit. DQM (and its variants) use a similar scheme
involving
both TTL and IPid.
However, they require
a lookup table to be stored in each router in advance, it
is a deterministic rather than a probabilistic scheme and requires two
bits in the IP
header. RAM like PCN is a probabilistic packet marking scheme. It also
uses a single ECN bit to mark packets, but it attempts to calculate
the load on the whole path rather than on each queue on the path
separately. It uses the IP TTL field to estimate the number of routers
in the network.

\subsection{Congestion estimation algorithm}
\label{sec:prediction}

Let $\rho_l$ be the load factor in time period
$l$ with
$l \in \{0,1,\ldots\}$.  Each time period is of length $t_\rho$
so $\rho_l$  is the estimate for time $[l t_\rho,(l+1) t_\rho)$.
The source estimates congestion every period $t_P$.
Consider the source attempting to estimate the load at the $i$th
intermediate router.
Let $e_l$ be the ratio of ECN marked ACKs to total
ACKs (markable by a given intermediate router $i$) in the $l$th such
period
$[l t_P, (l+1) t_P)$.  If this time period is
short then insufficient packets will be received to get a good
estimate.
If it is too long, the estimate will not capture the
dynamic nature of the traffic.

Let $L_l$ be the mean value of the load factor $\rho$ in the time
period
$[l t_P, (l+1) t_P)$.  Note that the load factor may have changed over
this period, particularly if $t_P$ (the time scale over which load
factor is estimated by the source) is significantly larger than
$t_\rho$
the time scale over which the load factor is held constant.  This
problem is to some extent unavoidable without time synchronisation
between source and router.
For the real traffic traces investigated in this work the congestion
level remained similar for
much longer time scales than $t_\rho$ and $t_P$.

One possible estimate for $L_{l+1}$ is $e_{l}$.
However, it is clear that if $L_l$ does not change too much between
given
time periods then
$e_{l-1}, e_{l-2}, \ldots$ can provide additional
information to help predict $L_{l+1}$. The approach taken here is to
use the well-known
Autoregressive Integrated Moving Average (ARIMA) models
to provide an improved predictor \cite{GBox}.  If
ARIMA models can be fitted to the time series $e_l$ this can be used
to get a good prediction of $L_{l+1}$
(since $e_l$ is an unbiased estimate of $L_l$ as previously proved).
The selection of an appropriate ARIMA model is discussed
in section \ref{sec:arimatune}.

\section{Simulation methodology}
\label{sec:method}
The protocol has been implemented using the simulation tool
ns-2\footnote{Network Simulator (ns-2), http://www.isi.edu/nsnam/ns}.
Real life traffic traces collected by CAIDA have been used to create
realistic background traffic (critically important for testing
the ARIMA model).  The simulation topology was chosen with the
following points in mind: 1)  The congestion levels should arise from
realistic traffic 2) The (artificially generated) PCN flows should
not significantly contribute to this congestion level 3)  The
congestion
at a given router should be a product of the mixing of more than
one traffic source.

\begin{figure*}[ht!]
\centering
\begin{center}
\includegraphics[scale=0.35]{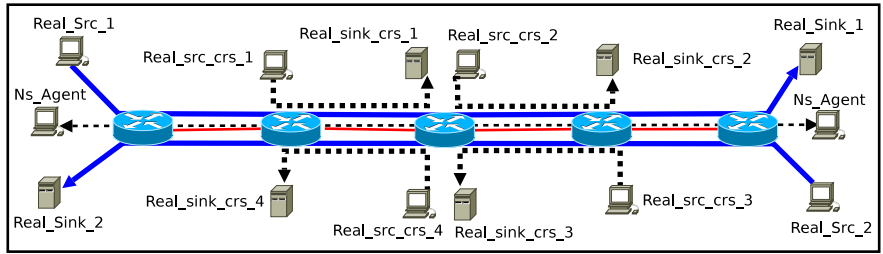}
\caption{Topology used for ns-2 simulation}
\label{fig:pklot}
\end{center}
\end{figure*}

The parking-lot topology used for the simulations is shown in figure
\ref{fig:pklot}.
The two hosts labelled {\tt Ns\_agent}, each sends PCN enabled traffic
to each
other.  For the purpose of congestion response in the simulation,
the traffic behaves exactly like TCP New-Reno but the
congestion marking from PCN is recorded.  This PCN traffic
crosses five intervening routers before reaching its destination.
Six flows derived from real traffic traces are used to provide
background
traffic.  These flows are between the hosts prefixed with {\tt Real}
on the diagram.

Two separate experiments are done using data derived from the CAIDA
project OC-12\footnote{OC12,
www.caida.org/data/passive/passive\_2007\_dataset.xml} and
OC-48\footnote{OC48,
www.caida.org/data/passive/passive\_oc48\_dataset.xml} traffic traces.
The traces known as
{\tt OC-12-1500} and {\tt OC-48-0900-0} are used, both hour long
traces.  In order to get six
real traces which have synchronised time behaviour, the CAIDA traces
were
each split by
source IP address into six separate traces. For the first
experiment the six data files from the OC-48 contained 79,550,409
packets
and those from the OC-12 contained 17,840,896 packets. This approach
is taken because these traces will then be correlated in time (traces
from 9am and 2am local time might exhibit very different traffic
behaviour).
Other traffic traces from these data sets have been tested with
similar
results to those reported in the next section.

UDP packets with length and inter-packet delay specified in these six
traces are
then fed into the network at each of the hosts suffixed {\tt
src\_$n$}.
Although obviously an ``open-loop" simulation like this does not
capture
the responsive nature of TCP, the authors consider it a more realistic
situation than using artificially generated TCP where the arrival
behaviour might be extremely unrealistic and hence unrealistically
easy
to predict.  This issue is, of course, extremely important when
considering
the utility of an ARIMA model.

Each router calculates its load factor at 200ms intervals, as
described in
(\ref{eqn:pcnlf}). The PCN hosts collect statistics at intervals of
length $t_P$
(this interval varies and $t_P > 200ms$) and calculate the proportion
of marked and unmarked packets for each intervening router which could
have
marked them.  This proportion provides an estimate of the mean load
factor
over the period $t_P$.  A better estimate for the mean load factor in
the next
time period can then be obtained using the time series forecasting as
described in section \ref{sec:prediction}. $t_P$ varies and the actual
load factors at each router and the predicted load factor at the
hosts is recorded for each of such intervals.

\section{Simulation results}
\label{sec:results}
In all the experiments in this section,
the first 10\% of the traffic is discarded as ns-2 warms up
and the second 10\% is used to estimate parameters for the time series
model.
As previously discussed, this load factor
changes every $t_{\rho}$ seconds (0.2 seconds).
Each PCN
source produces an estimate for congestion at
each outgoing queue every $t_P$ seconds.
The time period $t_P$ varies between 0.2 and 3.2 seconds.
This allows investigation of PCN behaviour as the number
of samples per time period changes. The predictor
$\hat{L_{i+1}}$ is compared with the crude estimate
obtained by assuming the
load factor to be $e_i$, the proportion of marked ACKs in the previous
time period.

\subsection{Determining the appropriate prediction model}
\label{sec:arimatune}

\begin{figure}[ht!]
\begin{center}
\includegraphics[width=5.6cm]{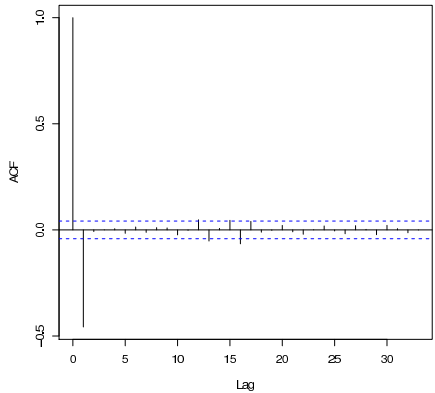}
\caption{The ACF of the differenced time series.}
\label{fig:acf}
\end{center}
\end{figure}

A general ARIMA($p,d,q$) model could be thought of as combining an
autoregressive AR($p$) component, a moving average MA($q$) component
and a differencing component $d$ -- for details see \cite{GBox}
amongst
many others.  The first task is
to determine the {\em order\/} of the model, that is
the values of $p, d$ and $q$.
For a given time series, in order to determine
the order of the appropriate ARIMA, the autocorrelation
function (ACF) and partial autocorrelation function (PACF), both
functions of lag $k$, are
examined.
If the ACF $\rho(k)$ becomes insignificant for $k > q$ for some small
$q$ then an AR($q$) model may be the best fit.  Similarly, if the PACF
$P(k$)
becomes insignificant for $k > p$ and some small $p$
then an AR($p$) model may be effective.
In the case of the $e_i$ time series examined here, in all
cases considered, the ACF of the differenced series $d_i = e_{i+1} -
e_i$
fell off quickly.  This can be seen in figure \ref{fig:acf} (the
dotted
lines indicate a 95\% confidence interval -- results between them are
likely to be merely noise) indicating only the lag one component is
very
significant (the lag zero component is by definition one).
This is an extremely strong indicator than an ARIMA(0,1,1) model is
appropriate.  Other ARIMA models were tested but
the ARIMA(0,1,1) proved to be the best.

\begin{figure}[ht!]
\begin{center}
\includegraphics[width=5.6cm]{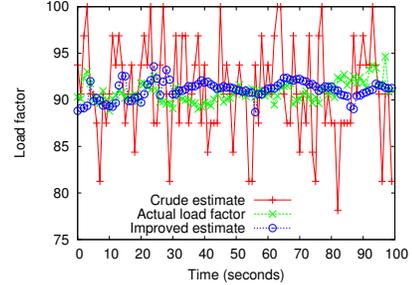}
\caption{An example of the ARIMA prediction.}
\label{fig:predict}
\end{center}
\end{figure}

A typical example of the prediction model running to improve the
raw estimates of load factor can be seen in figure
\ref{fig:predict}.
The figure plots the crude estimate and the improved estimate
against the actual load factor (arising from the real traffic) which
can only be observed at the router, not at the source.  In this
particular
sample the load factor remains relatively constant and high at around
90 for the whole period examined.
At almost all points in time the ARIMA modelling is closer to the real
load factor than the crude estimate.  More formal analysis of the
modelling error will be given in the next section.

\subsection{Estimation results}

The performance of $e_i$ and $\widehat{L_{i+1}}$ as estimators of
$L_{i+1}$
can be compared
against the bias of the estimate and the root mean square error (RMSE)
of
the prediction, which are for $N$ data points,
$\sum_i (L_i - \widehat{L_i})/N$ and $\sqrt { \sum_i (L_i -
\widehat{L_i})^2/N}$ respectively.  The RMSE would be expected to
change as a function of
$t_P$ -- as $t_P$ becomes smaller then $e_i$ is constructed from fewer
ACKs and hence would be expected to be a more inaccurate estimate of
$L_i$.
However, the situation is not quite as clear as that since,
if $t_P$ is small,
$L_{i+1}$ might be expected to be very
close in value to $L_i$ (since the load would not be expected to
change too much over small time periods).

In this section, the simulation described in section
\ref{sec:method} is run for values of $t_P$ from 0.2 to 3.2 seconds.
Each source produces estimates for the load $L_{i+1}$ using ARIMA
(0,1,1)
and the RMSE (and bias) are measured for the crude and corrected
estimators.

\begin{figure*}[ht!]
\begin{center}
\includegraphics[width=5.6cm]{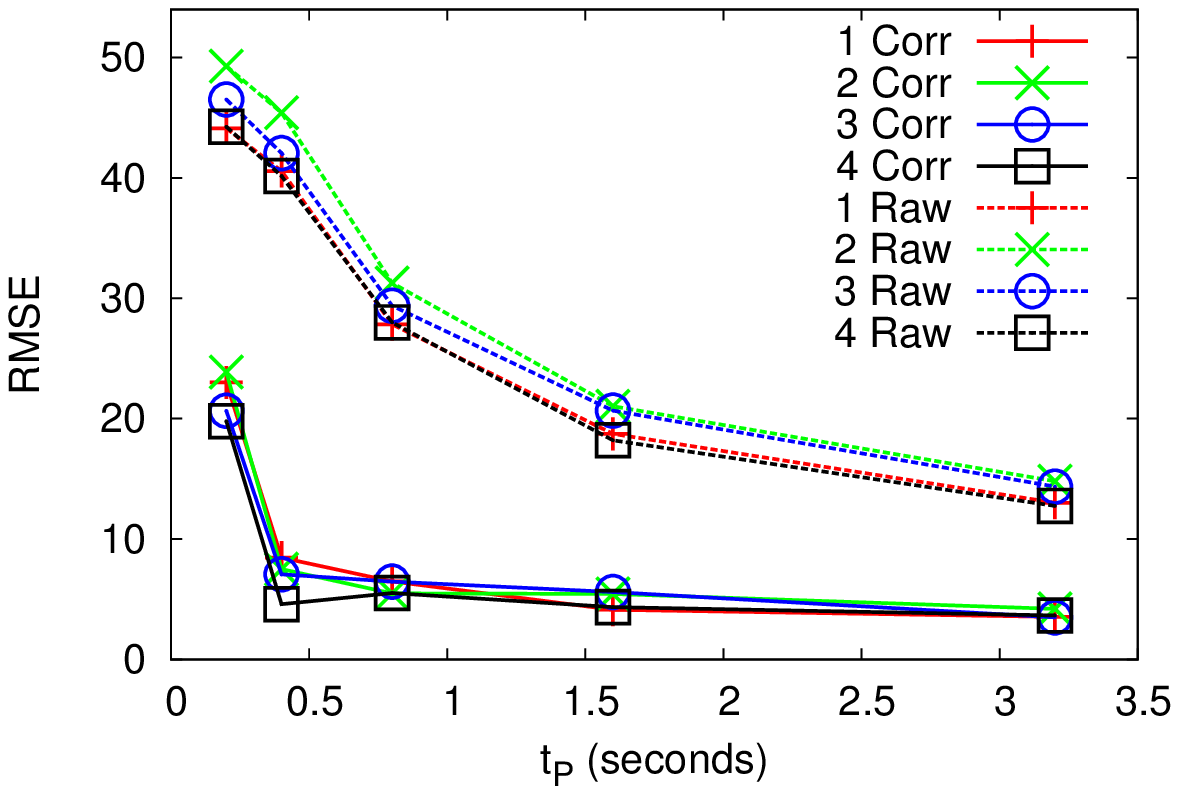}\includegraphics[width=5.6cm]{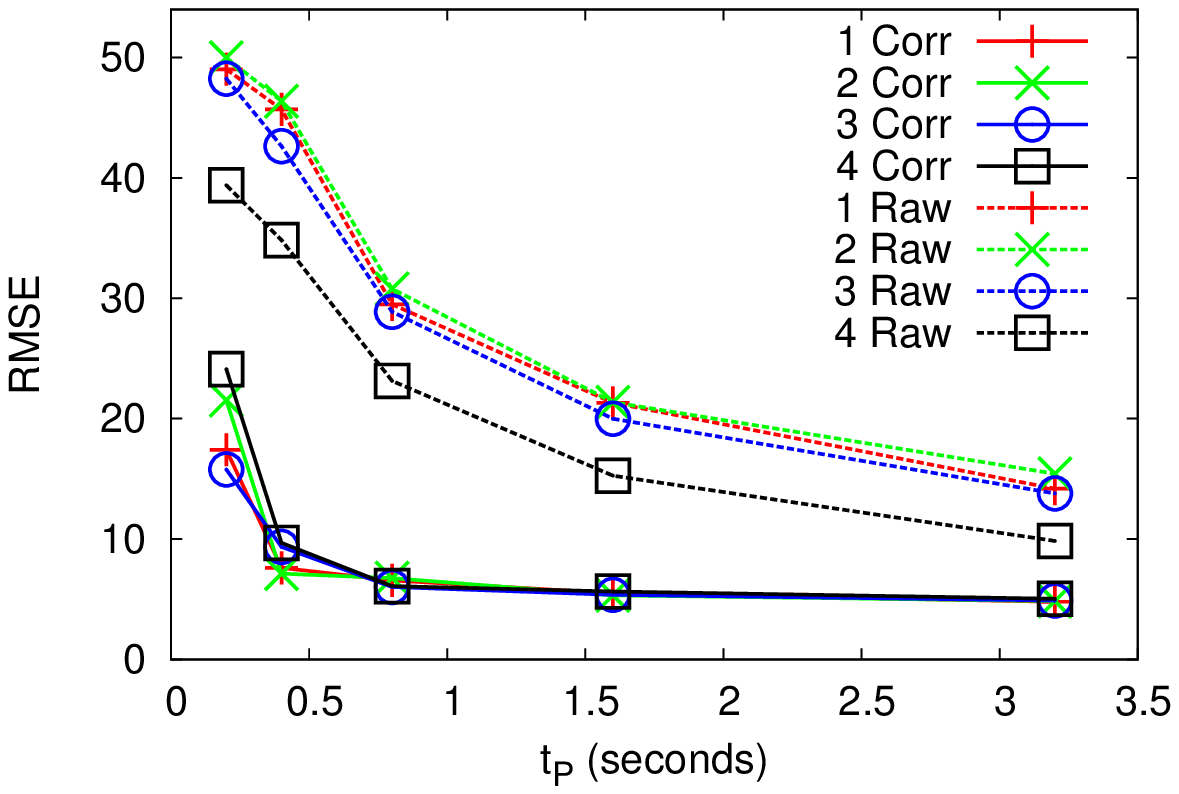}
\includegraphics[width=5.6cm]{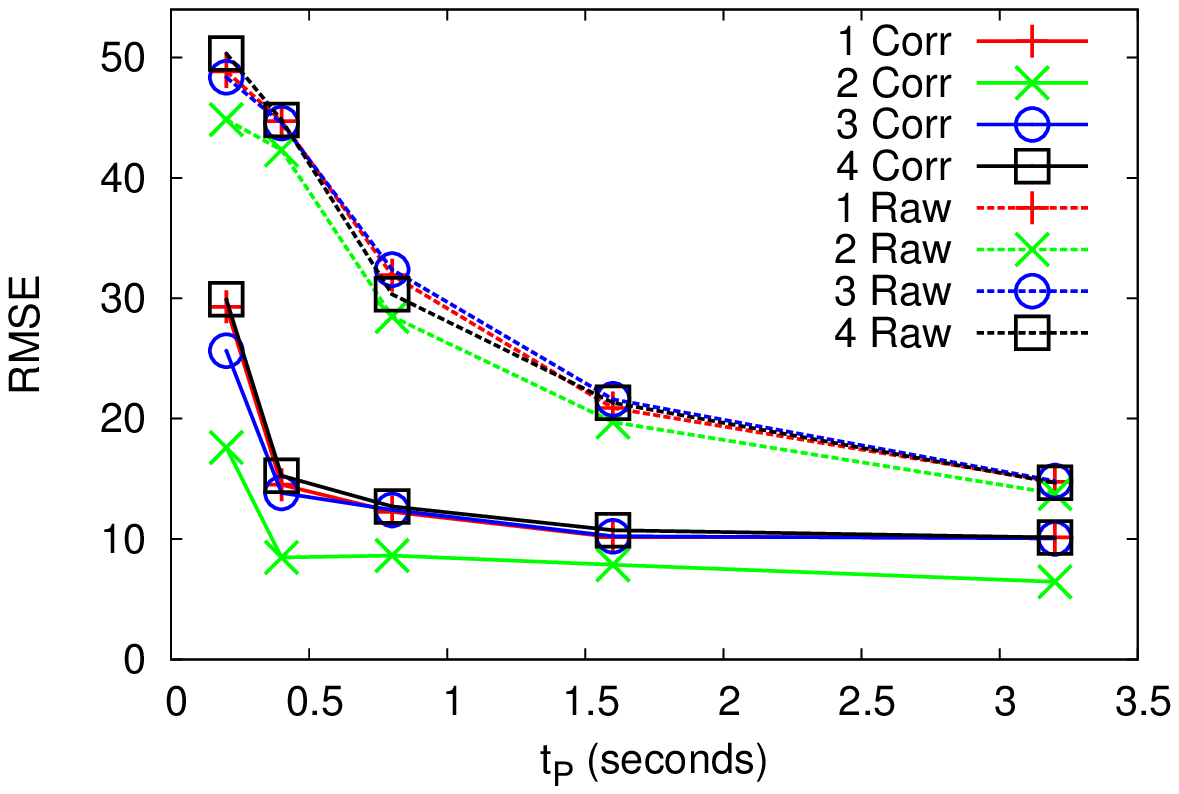}\includegraphics[width=5.6cm]{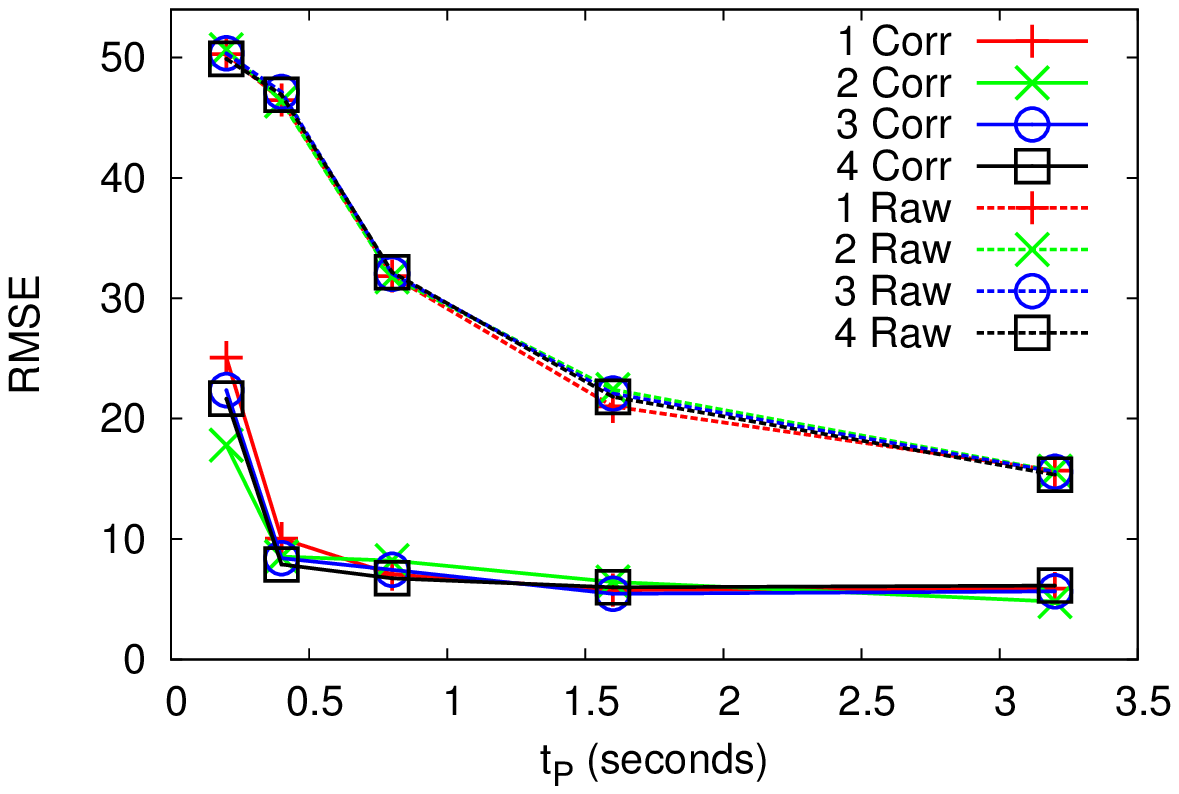}
\caption{Raw and corrected predictions RMSE for traffic originating
from the left source (left) and
the right source (right) for OC48 traffic (top) and OC12 traffic (bottom).}
\label{fig:rmse_all}
\end{center}
\end{figure*}

Figure \ref{fig:rmse_all} (top) shows for OC48 traffic the RMSE for 
the four queues outbound
(left top) and inbound (right top).  The dashed lines are the uncorrected
results $e_i$ and the solid lines are the results corrected by
the ARIMA process $\hat{L_i}$.
As can be seen, the longer the sampling period, the
better the estimate produced for the real load factor
(for both raw and ARIMA estimates).  In all cases, the ARIMA
procedure produced an extremely noticeable reduction in the error.
For
all but the shortest sampling period the RMSE was below 10 in the
corrected
data which, given the range is $[0,100]$ the method is producing an
extremely close prediction for the load factor in most sampled
periods.
Because of space constraints,
graphs for bias are not produced here but in all routers for all time
periods tested the bias was below five in magnitude and in the large
majority of cases below one in magnitude.  The conclusion is
that bias is not a significant problem.  As might be
expected, the ARIMA modelling (which proceeds from the $e_i$ anyway)
does
not correct the bias.

Figure \ref{fig:rmse_all} (bottom)
shows similar results for real traffic taken
from
the OC12 network and the topology in figure \ref{fig:pklot}.  The
pattern is the same as for
the OC48 traffic broadly speaking.  The main differences is that the
corrected RMSE is slightly larger for the data heading right.  For the
inbound
data the RMSE (particularly uncorrected)
is strikingly similar across all routers -- the small differences in
the data do not show up much in the plot.  (This is thought to
be the result of a single data stream being dominant in causing the
congestion inbound and hence similar results on all routers.)

Tests were also run on the pre-signalling idea which, by getting a
correct estimate of the number of routers on the path, greatly
improved
predictions.  This is partly a result of the shortness of the path
being tested here, five routers.  Pre-signalling therefore led to
an increase of more than six fold in the number of samples per
time period to get each $e_i$.  As would be expected the RMSE and
absolute value of bias were decreased greatly as a result.


\section{Conclusion}
\label{sec:conclusion}
PCN is a new method for determining the load on any of a number
(up to a fixed maximum) of intervening routers.  Related methods
exist in the literature but PCN has several advantages: 1)  It
produces an estimate of the load at each router. 2)  It estimates the
exact load rather than exactly determining whether the load is in one
of a small number of load regions. 3)  It requires the use of only one
ECN bit.

The raw prediction results for PCN have relatively high errors
in estimation of load factors but this can be corrected by using
the time-series technique ARIMA modelling.  This time series technique
could also be useful for a variety of other probabilistic packet
marking
schemes.

The obvious future work would be to use the predicted load factors
to produce an appropriate congestion response.  It would also be
of interest to implement the PCN protocol in a real kernel to see
how it performs in a real life situation.


\bibliographystyle{IEEEtran}
\bibliography{IEEEabrv,iccreferen}

\end{document}